\documentclass[aps,prl,reprint,showpacs,showkeys,unsortedaddress,superscriptaddress]{revtex4-1}
\usepackage{latexsym,amsmath,amssymb,bm}
\usepackage{graphicx,color}

\begin{document}

\title{Dynamics of colloidal aggregation in microgravity by critical Casimir forces}

\author{Marco A. C. Potenza}
\author{Andrea Manca}
\affiliation{Department of Physics, University of Milan, via Celoria 16,  I-20133 Milan, Italy}
\author{Sandra J. Veen}
\author{Bart Weber}
\affiliation{Van der Waals Zeeman Institute, University of Amsterdam, Science Park 904, 1098 XH Amsterdam, The Netherlands}
\author{Stefano Mazzoni}
\affiliation{European Space Agency/ESTEC, Keplerlaan 1, 2200AG Noordwijk, The Netherlands}
\author{Peter Schall}
\author{Gerard H. Wegdam}
\affiliation{Van der Waals Zeeman Institute, University of Amsterdam, Science Park 904, 1098 XH Amsterdam, The Netherlands}

\date{\today}

\begin{abstract}
Using the critical Casimir force, we study the attractive-strength dependence of diffusion-limited colloidal aggregation in microgravity. By means of near field scattering we measure both the static and dynamic structure factor of the aggregates as the aggregation process evolves. The simultaneous measurement of both the static and dynamic structure factor under ideal microgravity conditions allows us to uniquely determine the ratio of the hydrodynamic and gyration radius as a function of the fractal dimension of the aggregate, enabling us to elucidate the internal structure of the aggregates as a function of the interaction potential. We find that the mass is evenly distributed in all objects with fractal dimension ranging from 2.55 for a shallow to 1.75 for the deepest potential.
\end{abstract}

\maketitle

Since long colloidal aggregation processes have been recognized as a class of phenomena that can be described by very general rules, often referred to as the universality of colloidal aggregation. After the fundamentals introduced by Verwey and Overbeek \cite{Verwey_Overbeek}, the experimental discovery that colloidal clusters are endowed with fractal structures~\cite{Weitz84} triggered  a lot of work theoretically~\cite{vanDongen85}, by simulations~\cite{Meakin85} and experimentally~\cite{experimental} to describe aggregation phenomena that are ubiquitous in nature, foods, and many other consumer products. Colloidal aggregation is central to the formation of gels in systems with short-range attraction~\cite{Lu08}. However, most of our understanding comes from limiting cases of infinitely strong particle attraction, where particles stick irreversibly. The current understanding can be roughly summed up by distinguishing two different regimes: diffusion limited aggregation (DLA) and reaction limited aggregation (RLA), depending on the slowest phenomenon limiting the cluster growth~\cite{vanDongen85}.

One of the big challenges lies in understanding the internal structure of the aggregates. Especially at low attractive strength, where particles can detach and rearrange, the situation is not clear. In this case, the internal structure would give important insight into the aggregate growth process and the nature of aggregation for
a case that is most relevant for natural aggregation phenomena. As shown by Wiltzius~\cite{Wiltzius87}, this internal structure can be addressed by determining both the gyration radius, $R_g$, and the hydrodynamic radius, $R_h$, and their ratio, $\beta$,  that is directly related to the density-density correlation function of the aggregate. However, the challenge in the determination of $\beta$ is that it requires simultaneous measurement of both static and dynamic light scattering (SLS and DLS) for large objects, which is a difficult task. Typically, for such large clusters, the Brownian motion becomes very slow and dynamic measurements are overshadowed by large-scale convective motion and sedimentation, making the measurement of hydrodynamic radii prohibitively difficult.

Here, we study the internal structure of colloidal aggregates formed at low to high attractive strength. We exploit data obtained on the International Space Station (ISS), where there is no convection or sedimentation, and pure diffusive motion of the aggregates is guaranteed. We realize an effective attractive potential of controllable strength and range by employing critical Casimir forces~\cite{Bechinger08,Bonn09}. These attractive forces result from the confinement of critical solvent fluctuations between the particle surfaces; their strength is determined by the correlation length of the solvent and hence by temperature. This effect thus offers an effective potential that adjusts with temperature on a molecular time scale, allowing us to study the aggregation process as a function of the attractive potential. The application of a recently developed method of near field scattering (NFS) provided the unique opportunity to have simultaneously both SLS and DLS measurements at many angles~\cite{Mazzoni13} to determine both $R_g$ and $R_h$. This allowed us to elucidate the structure of the aggregates under ideal conditions and at finite potential. At low attractive strength, particles can be expected to detach and rearrange to form denser structures. Indeed, from the measured static form factors, the fractal dimension $d_f$ was observed to depend upon the interaction strength~\cite{Veen12}. Here, we show how the particle attraction influences the internal structure of the aggregates. We elucidate the internal structure and present experimental results for the dependence of the ratio $\beta=R_h/R_g$ upon the fractal dimension $d_f$. These measurements provide a deeper insight into the influence of the potential on the aggregation process.

The experiment, named COLLOID, operated in the ESA Mirogravity Science Glovebox, under the ELIPS program. Charge stabilized fluorinated latex particles 400 nm in diameter, with a density of 1.6 g/mL and refractive index of $n_p$ =1.37~\cite{Koenderink01} were suspended in a mixture of 3-methyl pyridine ($3MP$) in water/heavy water. Weight fractions of $X_{hw}$ = 0.63 for $D_2O/H_2O$ and $X_{3MP}$ = 0.39 for $3MP$ have been used, the solvent refractive index being 1.40. Before adding colloids, the solvent mixture was purified by distillation under vacuum. Four different suspensions were prepared, each one with a colloid volume fraction of $\sim10^{-4}$, and with different salt concentrations of 0.31, 1.5, 2.7 mmol/L of sodium chloride. The corresponding Debye screening lengths are, respectively, $L_D$= 14 nm, $L_D$= 6.4 nm, $L_D$= 4.8 nm. The samples were filled into quartz cells under vacuum that were tightly sealed.

We used a collimated laser beam 8~mm in diameter with a wavelength of 930 nm to illuminate the sample cell, and a 0.25 NA, 20X microscope objective to project the transmitted and scattered light onto a CCD sensor. As detailed in~\cite{Veen12}, near-field scattering images result from the interference between the intense transmitted and the (fainter) scattered light; these provide directly the scattered field-field correlation functions of the objects within the range of scattering angles selected by the lens.

We first determined the aggregation temperature, $T_{agg}$, of each sample using the second moment $m_2$ of the scattered intensity. The second moment is proportional to the cross-section and the number of scatterers; aggregation is detected when the second moment passes a threshold of $10^{-3}$. We then followed the entire aggregation process in time after temperature jumps from below $T_{agg}$ to $T_{agg}$, $T_{agg}$ + 0.1, + 0.2, + 0.3 and + 0.4$K$, increasing the attractive strength with each jump. For each temperature, we monitored the aggregation process for one hour by acquiring NFS images with a frame rate of 1 $s^{-1}$ in batches of 100. The sample was then cooled down to a temperature far below $T_{agg}$ to split up the aggregates, followed by stirring for at least 3 hours before a new measurement was started.

\begin{figure}
	\includegraphics[width=1.0\columnwidth]{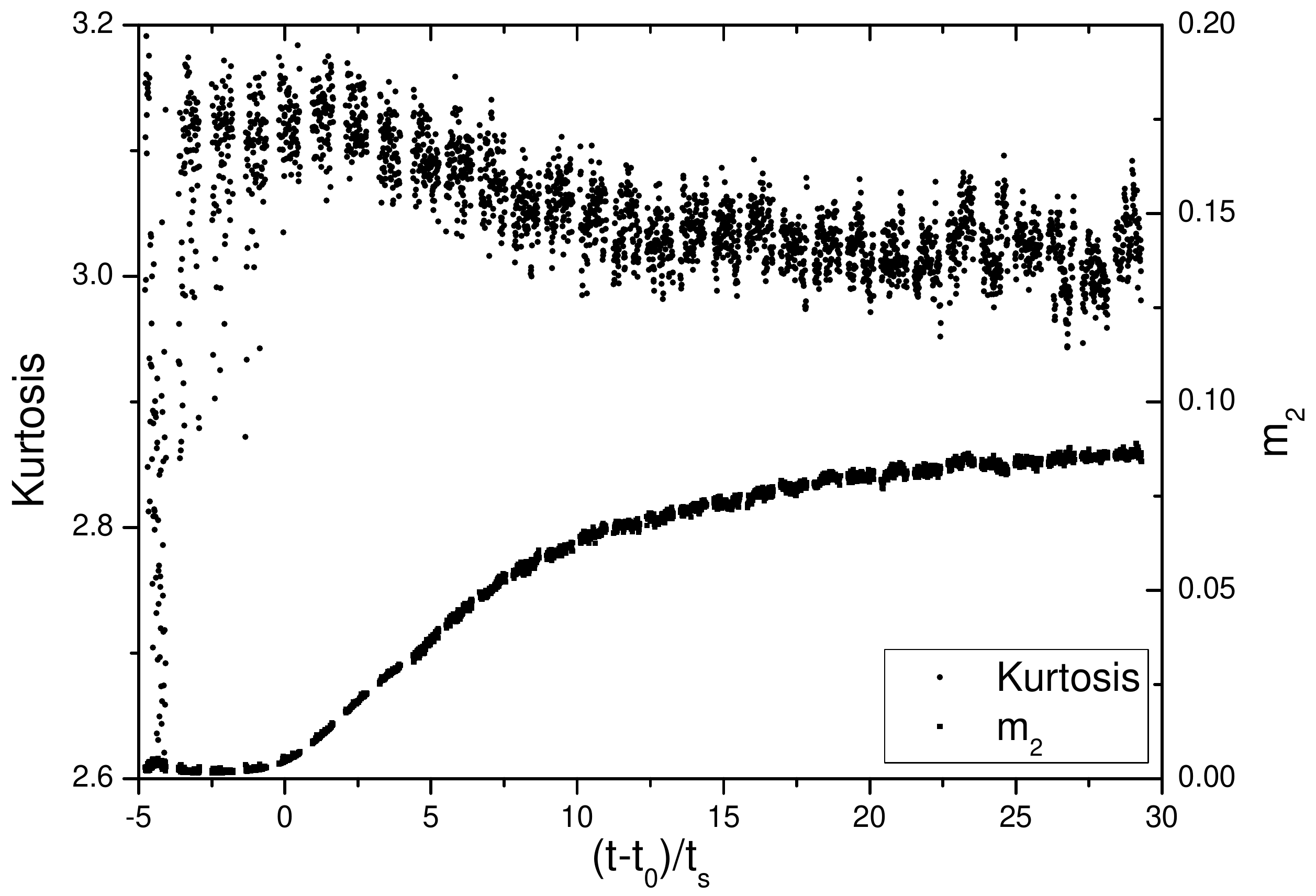}
    \caption{Evolution of the second moment and the kurtosis during the aggregation process at the lowest attractive strength at $T = T_{agg}$.}
	\label{fig1}
\end{figure}

\begin{figure}
	\includegraphics[width=1.0\columnwidth]{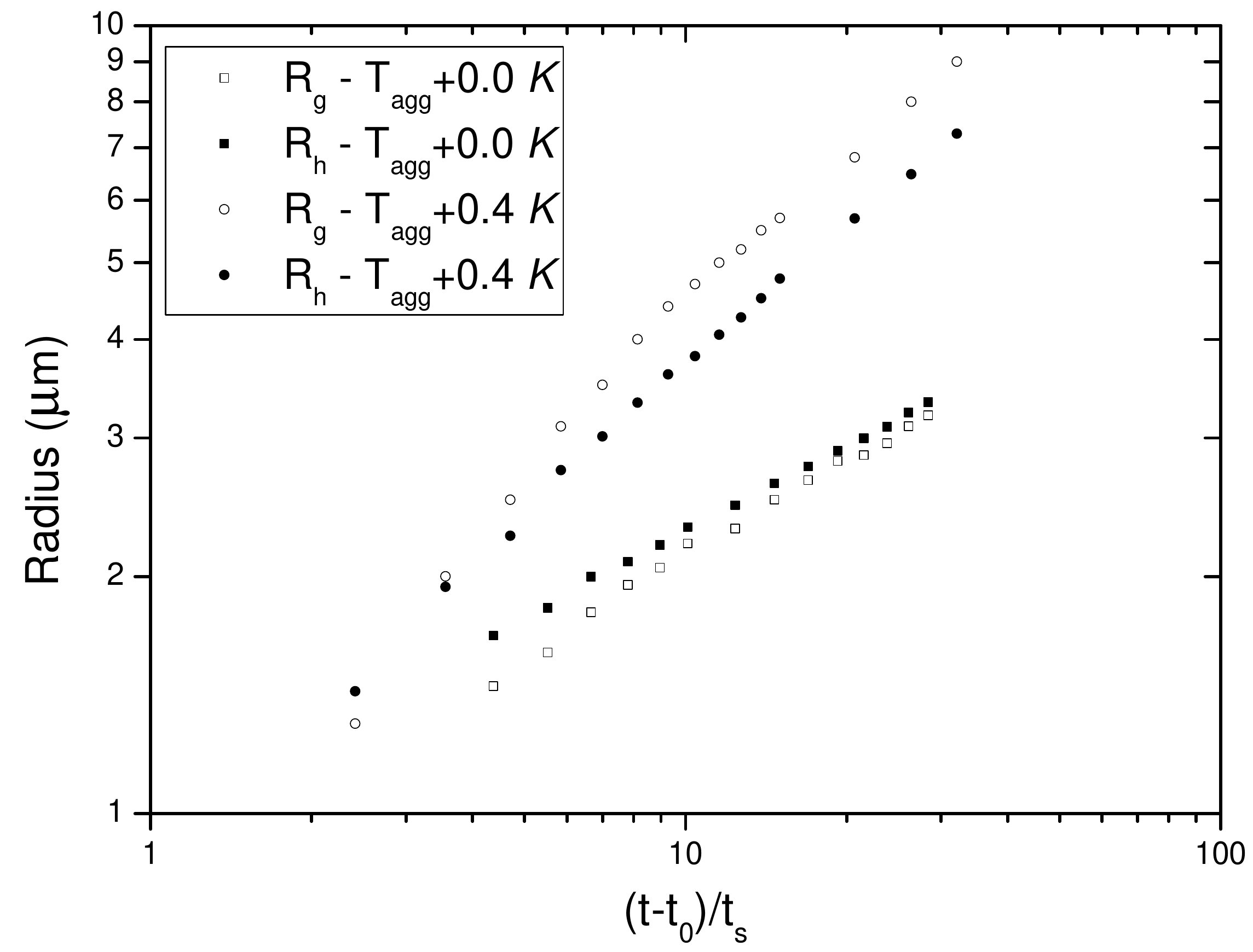}
	\caption{Aggregate growth: Evolution of $R_g$ and $R_h$ with time. The radius of gyration was obtained from the Fisher-Burford fit of the static power spectra. The hydrodynamic radius was obtained using eq.~\ref{eq1}.}
	\label{fig2}
\end{figure}

A typical example of the evolution of $m_2$ is shown in Fig.~\ref{fig1}. Here, we have used the reduced time $t_r = (t-t_0)/t_s$, where $t_s$ is the time a single particle diffuses over its own diameter and $t_0$ is the start of the aggregation process, which we define by linear extrapolation of the data to vanishing $m_2$. Starting from $t_0$, the second moment rises continuously, indicating the growth of aggregates. In this regime, the $m_2$ curves of all samples at all temperatures can be scaled onto a single master curve, reflecting the universal growth of the aggregates. As a guide and measure of reliability, we also indicate the kurtosis, $\kappa = m_4/m_2^2$. The kurtosis varies wildly for $t < t_0$ and reaches a defined value of about 3 after that. We hypothesize that at these early times before $t_0$, subcritical nuclei may form and evaporate on the time scale of observation. After this, we can measure a reliable radius and diffusion coefficient of the scattering objects.

Microgravity conditions allow unique measurement of the internal structure of the aggregates: they permit the slow Brownian motion of large aggregates to be measured, until the aggregates exhibit the pronounced form factors of fractal objects. The NFS measurement technique then allows us to measure the intermediate scattering function $F(q,\tau)$ instantaneously with respect to the much slower aggregation process and diffusion time, enabling us to determine the evolution of both the hydrodynamic and the gyration radius as the aggregates grow.

We first determine the gyration radius from the static form factor $S(q)$. This is done using the Fisher-Burford fit $S(q,R_g) = \left(1+(2/3d_f)q^2R_g^2 \right)^{-d_f/2}$ for fractal aggregates that depends only on the radius of gyration and the fractal dimension, $d_f$~\cite{Veen12}. The resulting evolution of $R_g$ is shown by open symbols in Fig.~\ref{fig2}. In good approximation, $R_g$ grows as a power law,  $R_g \sim t_r^{1/d_f}$, as expected for a pure DLA process. In pure DLA, the inverse exponent equals the fractal dimension of the aggregate. We find that, indeed, the slopes in Fig.~\ref{fig2} are in good agreement with the fractal dimension $d_f=2.4$ and $d_f=1.8$ determined from the static structure factor respectively at $T_{agg}$ and $T_{agg}+0.4K$~\cite{Veen12}. Hence, we can describe the entire aggregation processes within the framework of DLA with one consistent value of the fractal dimension.

To determine the hydrodynamic radius, we relate the decay of the intermediate scattering function $F(q,t)$ to the diffusion coefficient, $D$. In the case of monodisperse simple spherical objects, $D$ is related to the hydrodynamic radius, $R_h$, through the Stokes-Einstein relation $D=\frac{k_bT}{6\pi \eta R_h}$, in which $\eta$ is the solvent viscosity. No dependence of $R_h$ on $q$ is expected. For our aggregates, however, $R_h$ as obtained from the decay of $F(q,t)$ varies with $q$: the additional rotational degrees of freedom of objects with internal structure lead to a faster decay of $F(q,t)$. In this case, the effective diffusion coefficient $D_{eff}$ as defined from the decay time $\tau$ via $\tau=1/D_{eff}q^2$ is no longer related to the hydrodynamic radius via the Stokes-Einstein relation. To account for the rotations quantitatively, we define a class of spherical symmetries that determines how much an object needs to rotate before it decorrelates. The higher the symmetry, the smaller is the angle and thus the time needed to decorrelate the field~\cite{Lattuada04}. Because the symmetry is connected to $qR_g$ via the scale-invariant fractal structure, $qR_g$ determines uniquely the correction to the decorrelation times. This allows us to relate $D_{eff}$ to $D$ via the static structure factor $S(q)$~\cite{Lin90}. For $S(q)$, we use the structure factor as measured at the same time. To be independent of fluctuations in $S(q)$, we again use the Fisher-Burford fit as detailed above; this expression provides a good fit to the measured average structure factor of the aggregates. Using it, we can rewrite eq. 7 from Lin et al.~\cite{Lin90} into the following form:

\begin{equation}
2\left(\frac{R_h}{R_g}\right)^{2}\left(\frac{D_{eff}}{D(R_h)}-1\right) = 1-\frac{3d_{f}}{3d_{f}+2(qR_{g})^{2}},
\label{eq1}
\end{equation}

where we have separated static and dynamic quantities. The right-hand side is given by the static data; it provides a master curve solely dependent on the product $qR_g$, taking into account both translations and rotations of the aggregates. The left-hand side is determined from the dynamics: taking $R_g$ from the static data, the only free parameter is the hydrodynamic radius. To determine it, we define $f(R_h)$ that equals the right and left-hand side of the equation, and determine diffusion coefficients $D_{eff}$ from the time decay of $F$ for 60 different $q$-values in the range 0.8 - 1.9 $\mu m^{-1}$. We then adjust $R_h$ so that $f(R_h)$ from the dynamic data provides the best fit to the master curve.

\begin{figure}
	\includegraphics[width=1.0\columnwidth]{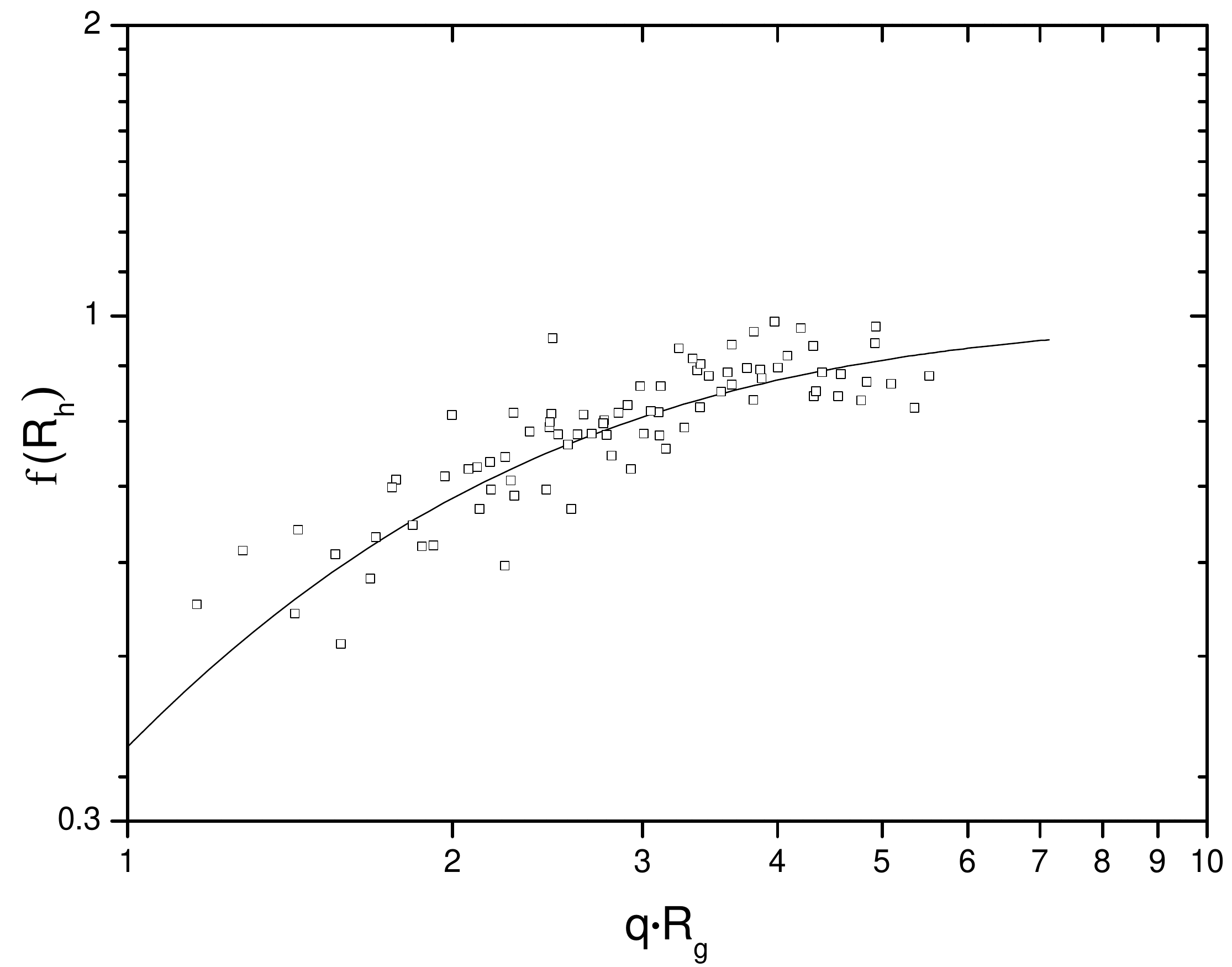}
	\caption{Master curve according to eq.~\ref{eq1} of the scattering of aggregates growing by diffusion-limited aggregation at the lowest attractive strength at $T = T_{agg}$.}
	\label{fig3}
\end{figure}

An example of the dynamic data compared to the master curve is shown in Fig.~\ref{fig3}. The weakest interaction is considered, namely at the temperature $T=T_{agg}$. The data is well fitted with the ratio $R_h/R_g$ = 1.05, one consistent value for the entire aggregation process. Fits of similar quality are obtained for all other interaction potentials as long as the aggregates are not too large. The situation changes for the highest interaction potential, when aggregates grow to larger sizes: the deviation from the master curve becomes larger due to polydispersity effects~\cite{Wiltzius87}. By taking the polydispersity into account we can reduce the deviations and again obtain a good fit for one value of $\beta$. We accounted for polydispersity explicitly assuming a cluster mass distribution of the form:

\begin{equation}
N(M)=\frac{N_T}{\left\langle M\right\rangle}\left[1-\frac{1}{\left\langle M\right\rangle}\right]^{M-1}
\label{eq2}
\end{equation}
where $N(M)$ is the number of clusters of mass $M$, $N_{T}$ is the total number of clusters, and $\langle M \rangle$ is the average cluster mass, which is related to the radius of gyration $R_g$ via $\langle M \rangle=(R_{g}/a)^{d_{f}}$, where $a$ is the monomer radius. This distribution represents a good description for the DLA case (see~\cite{Lin90} and references therein for details). To determine the effective diffusion coefficient, for each cluster mass, we determined the individual diffusion coefficients using eq. \ref{eq1}, and we averaged the diffusion coefficients of all clusters, taking into account their statistical weight given by $N(M)$~\cite{Lin90,Berne_Percora}. The results of this method agree with the earlier one that neglects polydispersity to within 2-3\% for small values of $qR_g$; however, by incorporating polydispersity, we now obtain a similarly accurate fit over the entire range of $qR_g$, even for large values of $qR_g$. We note that this large-aggregate regime is important in microgravity measurements such as the one explored here, because  large aggregates do not settle on the observation time scale and stay in the field of view. In contrast, this regime is rarely approached on earth, and is not reached in~\cite{Lin90} because of the fast settling of large aggregates.

Thus, we have obtained the hydrodynamic radius upon growth of the aggregates for a large range of sizes. The resulting values of $R_h$ as a function of time are indicated by closed symbols in Fig.~\ref{fig2}. Comparison with $R_g$ in the same figure shows that the ratio $R_h/R_g$ becomes constant for larger aggregates, being $\beta=1.1\pm0.012$ for $T=T_{agg}$, and $\beta=0.85\pm0.015$ for $T=T_{agg}+0.4 K$. For smaller aggregates, the ratio $R_h/R_g$ appears not constant because the aggregates cannot be considered spherical as the monomers are 0.4 $\mu m$ in diameter and therefore aggregates consist of a few monomers only. This small size effect is also observed in the growth as a deviation from the expected power law $R_g=at^{1/d_{f}}$ dependence. By considering only sufficiently large aggregates that have a well-defined fractal dimension and structure, we can now elucidate the internal structure as a function of the attractive potential and fractal dimension.

\begin{figure}
	\includegraphics[width=1.0\columnwidth]{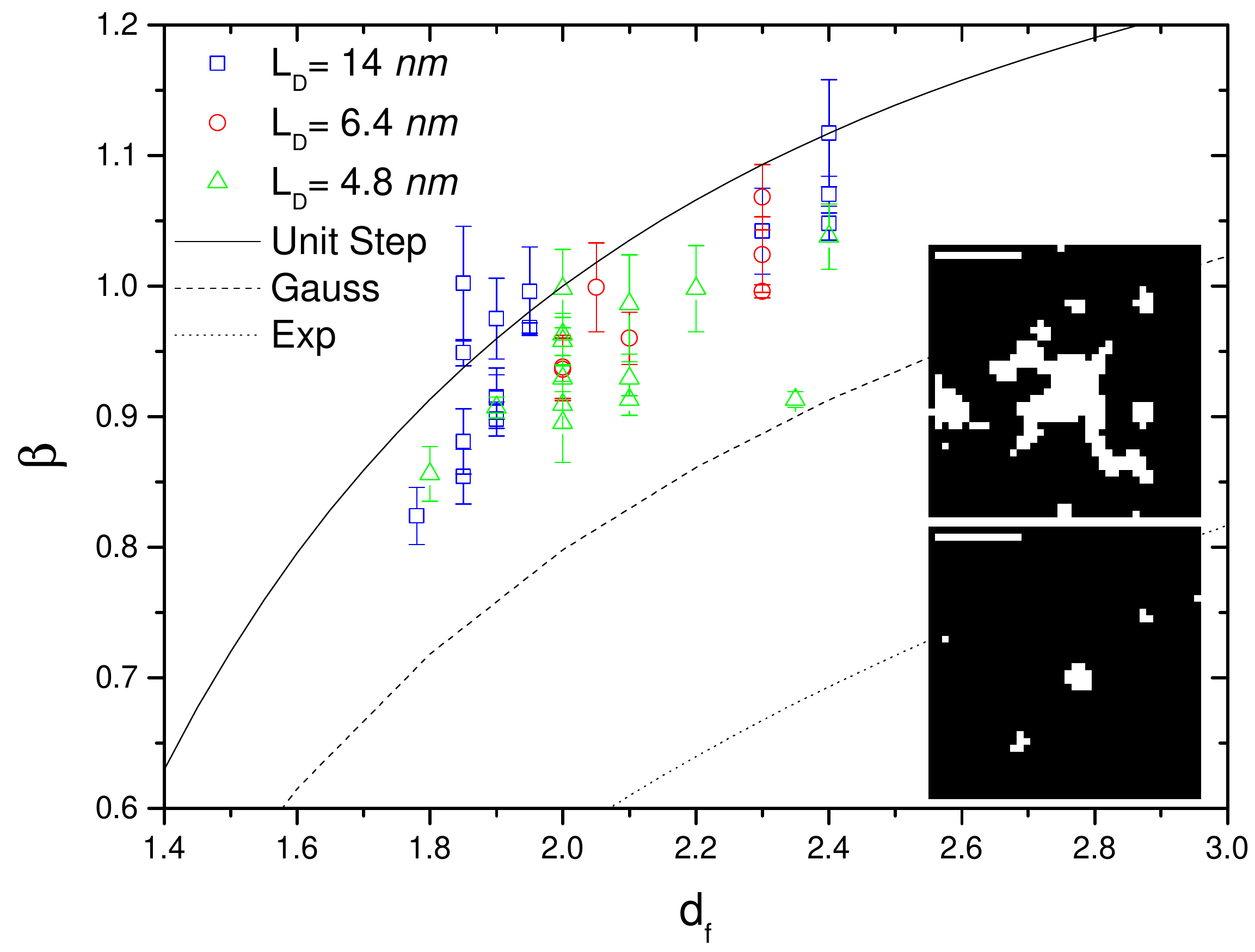}
	\caption{Ratio of hydrodynamic to gyration radius as a function of $d_f$ for three different salt concentrations (see text for details). Lines indicate the expected dependence obtained for unit step, Gaussian and exponential density-density correlation functions (from top to bottom). Insets show holographic reconstructions of aggregates grown at $T = T_{agg} + 0.4K$ (highest attraction, top) and $T = T_{agg}$ (lowest attraction, bottom). The length of the scale bar is 25$\mu$m.}
	\label{fig4}
\end{figure}

We plot the ratio $\beta = R_h / R_g$ as a function of $d_f$ in Fig.~\ref{fig4}. Measurements at all temperatures from $T_{agg}$ to $T_{agg}+0.4 K$, and for three different salt concentrations (0.31mmol/L, blue squares; 1.5 mmol/L, red circles; 2.7mmol/L, green triangles) are included. A systematic dependence of $\beta$ on $d_f$ is observed, indicating a consistent internal structure independent of the salt concentration. To interpret the $\beta$-values, we indicate by lines the dependence for a fully compact object with a unit-step density distribution (solid curve), for a fluffier object with a Gaussian density profile (dashed curve), and for an exponential profile (dotted curve)~\cite{Wiltzius87}. The data lies closest to the unit step, indicating that the aggregates have fairly compact internal structure, regardless of the fractal dimension and thus the attractive strength. To illustrate the shape of the aggregates, as insets we show results obtained by holographic reconstruction of the aggregates at the latest stages of aggregation for the two conditions considered in Fig.~\ref{fig2}. The fluffier structure of aggregates formed at higher attractive interaction strength is clearly observable.

In summary, we measured the internal structure of DLA clusters over a wide range of attractive potential strength. This was possible owing to the peculiarity of the critical Casimir effect allowing us to induce tunable interactions that result in aggregates with a wide range of fractal dimensions. Notice that no systematic dependence on the added salt is observed, and the only dependence is on the fractal dimension $d_f$, set by the depth of the potential well~\cite{Veen12}. The salt concentration i.e. the repulsive part of the potential has no effect on $d_f$, as it is only the depth of the potential well that determines the fractal dimension. While the aggregation temperature is set by the ratio of the solvent correlation length to the Debye screening length, it is the temperature increment from $T_{agg}$ that sets the depth of the potential well, and consequently determines the aggregate structure, i.e. its fractal dimension. This is the first time that a DLA process is studied with the widest range of possible fractal structures, owing to the variable critical Casimir potential, microgravity and the near field scattering technique that measures the intermediate structure function as a function of time.


We thank Matteo Alaimo for his support with the analysis code. This work was supported by the European Space Agency and the Dutch organization for scientific research NWO.


\begin{thebibliography}{99}


\bibitem{Verwey_Overbeek}
E. J. W. Verwey and J. T. G. Overbeek, \emph{Theory of the Stability of Lyophobic Colloids}, Elsevier, Amsterdam (1948).

\bibitem{Weitz84}
D.A. Weitz and M. Oliveira, \emph{Phys. Rev. Lett.} {\bf 52}, 1433 (1984); C. Aubert and D.S. Cannell, \emph{Phys. Rev. Lett.} {\bf 56}, 738 (1987).

\bibitem{vanDongen85}
G. J. van Dongen and M. H. Ernst, \emph{Phys. Rev. Lett.} {\bf 54}, 1396 (1985).
	
\bibitem{Meakin85}
P. Meakin, T. Vicsek, and F. Family, \emph{Phys. Rev. B} {\bf 31}, 564 (1985)

\bibitem{experimental}
D. A. Weitz, J. S. Huang, M.Y. Lin, and J. Sung, \emph{Phys. Rev. Lett.} {\bf 54}, 1416 (1985); R. C. Ball, D. A.Weitz, T. A.Witten, and F. Leyvraz, \emph{Phys. Rev. Lett.} {\bf 58}, 274 (1987); M.Y. Lin, H. M. Lindsay, D.A. Weitz, R. C. Ball, R. Klein, and P. Meakin, \emph{Nature} (London) {\bf 339}, 360 (1989).

\bibitem{Lu08}
P.J. Lu, Emanuela Zaccarelli, Fabio Ciulla, Andrew B. Scofield, Francesco Sciortino, and D.A. Weitz, \emph{Nature}  {\bf 453}, 499 (2008)

\bibitem{Wiltzius87}
P. Wiltzius, \emph{Phys. Rev. Lett.} {\bf 58}, 710 (1987); P. Pusey et al., \emph{Phys. Rev. Lett.} {\bf 59}, 2122 (1987); Wiltzius and Saarloos, \emph{Phys. Rev. Lett.} {\bf 59}, 2123 (1987)

\bibitem{Bechinger08}
C. Hertlein, L. Helden, A. Gambassi, S. Dietrich, C. Bechinger, \emph{Nature}  {\bf 451}, 172 (2008).

\bibitem{Bonn09}
D. Bonn, J. Otwinowski, S. Sacanna, H. Guo, G. Wegdam, and P.Schall, \emph{Phys. Rev. Lett.} {\bf 103}, 156101 (2009).

\bibitem{Mazzoni13}
S. Mazzoni., et al., \emph{Rev. Sci. Instrum.}, in press (2013).

\bibitem{Veen12}
S. Veen, et al, \emph{Phys. Rev. Lett.} {\bf 109}, 248302 (2012)

\bibitem{Koenderink01}
G. H. Koenderink, S. Sacanna, C. Pathmamanoharan, M. Ras, and A. P. Philipse, \emph{Langmuir} {\bf 17}, 6086 (2001).

\bibitem{Lattuada04}
M. Lattuada, H. Wu, M. Morbidelli, \emph{Langmuir} {\bf 20}, 5630 (2004)

\bibitem{Lin90}
M.Y. Lin, H.M. Lindsay, D.A. Weitz, R. Klein, R.C. Ball, and P. Meakin, \emph{Phys. Rev. A} {\bf 41}, 2005 (1990)

\bibitem{Berne_Percora}
B.J. Berne, R. Pecora, \emph{Dynamic Light Scattering}, Dover NY 2000


\end{thebibliography}
\end{document}